\documentclass[11pt]{article} 
\usepackage{amsmath}
\usepackage{amssymb}
\usepackage{amsthm,amsxtra}
\usepackage{epsf}
\usepackage{eepic}
\newlength{\mytopmargin}
\newlength{\myleftmargin}
\newtheorem{thm}{Theorem}
\newtheorem{cor}{Corollary}

\newtheorem{prop}{Proposition}
\begin{document}
\vspace{4cm}
\begin{center}
{\Large \bf Hard and soft edge spacing distributions for random matrix ensembles with
orthogonal and symplectic symmetry} 
\end{center}

\vspace{5mm}
\begin{center}
P.~J.~Forrester \\
Department of Mathematics and Statistics,
University of Melbourne, 
Victoria 3010, Australia 
\end{center}

\small
\begin{quote}
Inter-relations between random matrix ensembles with different symmetry types provide
inter-relations between generating functions for the gap probabilites at the spectrum edge.
Combining these in the scaled limit with the exact evaluation of the gap probabilities for
certain superimposed ensembles with orthogonal symmetry allows for the exact evaluation of
the gap probabilities at the hard and soft spectrum edges in the cases of orthogonal and symplectic symmetry.
These exact evaluations are given in terms of Painlev\'e transcendents, and in terms of
Fredholm determinants.
\end{quote}

\section{Introduction}
\subsection{Background}
There are many spacing distributions in random matrix theory which can be expressed in terms
of Painlev\'e transcendents 
(see \cite{Fo05a} for
a recent review and \cite{Fo02} for an extended treatment).
How this comes about in the case of random
matrix ensembles with a unitary symmetry is easy to explain. Thus in this
setting the underlying joint eigenvalue probability density function (PDF) is
of the form
\begin{equation}\label{1}
{1 \over C} \prod_{l=1}^N g(x_l) \prod_{1 \le j < k \le N} (x_k - x_j)^2 =:
{\rm UE}_N(g),
\end{equation}
where $C$ denotes the normalization. Spacing distributions relating to an interval $J$
can be computed from the generating function
\begin{equation}\label{2}
\Big \langle \prod_{l=1}^N (1 - \xi \chi_J^{(l)} )\Big \rangle_{{\rm UE}_N(g)},
\end{equation}
where $\chi_J^{(l)} = 1$ for $x_l \in J$, and $\chi_J^{(l)} = 0$ otherwise. Using a well
known identity due to Heine the multidimensional integral (\ref{2}) can be written as
a determinant. In the case of $g(x)$ a classical weight function (see \cite{AFNV00}
for a precise definition), this determinant can be written in a double Wronskian form.
As such it is recognised as Sylvester's solution of the Toda lattice equation
appearing in  Okamoto's Hamiltonian formulation of the Painlev\'e equations \cite{Ok87b}.

In the case of spacing distributions in random matrix ensembles with orthogonal or
symplectic symmetry, the relationship with Painlev\'e theory requires further insights.
(We remark that the symmetry corresponds to the subgroup of unitary matrices which
diagonalise the matrices in the corresponding ensemble.) Such insight
comes by way of certain inter-relationships amongst random matrix ensembles with
the different symmetry types.

First we recall that the joint  eigenvalue PDF is
\begin{equation}\label{3}
{1 \over C} \prod_{l=1}^N g(x_l) \prod_{1 \le j < k \le N} |x_k - x_j|^\beta =:
\left \{ \begin{array}{ll}{\rm OE}_N(g), & \beta = 1 \\
{\rm SE}_N(g), & \beta = 4 \end{array} \right.
\end{equation}
in the case of orthogonal and symplectic symmetry respectively. The definitions (\ref{1}) and
(\ref{3}) assume the eigenvalues are all real (Hermitian matrices). Historically
inter-relations between matrix ensembles with different symmetry types were first formulated
for Dyson's circular ensemble of random matrices. These are ensembles of unitary matrices, and
so the eigenvalues are on the unit circle in the complex plane. The joint eigenvalue PDF
is
\begin{equation}\label{4}
{1 \over C} \prod_{1 \le j < k \le N} |e^{i \theta_k} - e^{i \theta_k} |^\beta =:
\left \{ \begin{array}{lll}{\rm COE}_N, & \beta = 1 \\
{\rm CUE}_N, & \beta = 2 \\
{\rm CSE}_N, & \beta = 4  \end{array} \right.
\end{equation}
where $\beta = 1,2$ or 4 according to the symmetry being orthogonal, unitary or
symplectic respectively.

The inter-relations can be derived from certain superposition and decimation procedures.
In relation to the former, consider two independent eigenvalue sequences drawn from COE${}_N$.
Let the corresponding eigenvalues have angles $0 < \phi_1 < \cdots < \phi_N < 2 \pi$ and
$0 < \phi_1' < \cdots < \phi_N' < 2 \pi$ respectively. Now let the sequences be
superimposed to form ${\rm COE}_N \cup {\rm COE}_N$, relabel the eigenvalues
$$
0 < \theta_1 < \theta_2 < \cdots < \theta_{2N} < 2 \pi
$$
and integrate over every second eigenvalue (e.g.~if the odd labelled sequence is to be
integrated over, integrate $\theta_{2j-1}$ over $(\theta_{2j-2}, \theta_{2j})$, where
$\theta_0:=\theta_{2N} - 2 \pi$. With this operation denoted alt, it was conjectured by
Dyson \cite{Dy62c}, and subsequently proved by Gunson \cite{Gu62}, that
\begin{equation}\label{C1}
{\rm alt} ( {\rm COE}_N \cup {\rm COE}_N ) =  {\rm CUE}_N.
\end{equation} 
The decimation procedure is to take an eigenvalue sequence from  COE${}_{2N}$, and to
integrate over every second eigenvalue. For this it was proved by Dyson and Mehta
\cite{DM63} that
\begin{equation}\label{C2}
{\rm alt}(  {\rm COE}_{2N} ) =  {\rm CSE}_N.
\end{equation}

The inter-relationships (\ref{C1})--(\ref{C2}) form the basis of corresponding
inter-relationships between generating functions for spacing distributions. Let
$E(n;J;{\rm ME}_N)$ denote the probability that in the matrix ensemble
${\rm ME}_N$ the interval $J$ contains exactly $n$ eigenvalues. Introduce the
generating functions 
\begin{eqnarray*}
&& E^{{\rm CUE}_N}((-\theta,\theta);\xi) :=
\sum_{n=0}^\infty (1 - \xi)^n E(n;(-\theta,\theta);{\rm CUE}_N) \\
&&E^{{\rm CSE}_N}((-\theta,\theta);\xi) :=
\sum_{n=0}^\infty (1 - \xi)^n E(n;(-\theta,\theta);{\rm CSE}_N) \\
 &&E^{{\rm COE}_N \pm}((-\theta,\theta);\xi) :=
\sum_{n=0}^\infty (1 - \xi)^n \Big ( E(2n;(-\theta,\theta);{\rm COE}_N) +
 E(2n\mp 1;(-\theta,\theta);{\rm COE}_N) \Big ).
\end{eqnarray*}
Then it follows from (\ref{C1}) and (\ref{C2}) that 
\begin{eqnarray}
&& E^{{\rm CUE}_N}((-\theta,\theta);\xi) =
E^{{\rm COE}+}((-\theta,\theta);\xi)E^{{\rm COE}-}((-\theta,\theta);\xi) \label{1.7w}\\
&& E^{{\rm CSE}_N}((-\theta,\theta);\xi) = {1 \over 2} \Big (
E^{{\rm COE}+}((-\theta,\theta);\xi) + E^{{\rm COE}-}((-\theta,\theta);\xi)
\Big ). \label{1.7x}
\end{eqnarray}

Analogous to the situation with (\ref{2}), the generating function $E^{{\rm CUE}_N}
((-\theta,\theta);\xi)$ can readily be expressed as a determinant. The determinant
can, in turn, be identified as the $\tau$-function of a certain Painlev\'e VI
system \cite{FW04}, and thus be characterised in terms of  Painlev\'e VI transcendents.
For the equations (\ref{1.7w}), (\ref{1.7x})  to similarly allow for the determination of
$\{E(n;(-\theta,\theta);{\rm COE}_N)\}$ and $\{E(n;(-\theta,\theta);{\rm CSE}_N)\}$ it is
necessary to evaluate one of $ E^{{\rm COE}_N \pm}$. In fact it is possible to deduce
that \cite{Fo06}
\begin{eqnarray}
 E^{{\rm COE}_N +}((-\theta, \theta); \xi) & = &
\Big \langle \prod_{l=1}^N (1 - \xi \chi_{(0,\sin^2 \theta/2)}^{(l)} ) \Big \rangle
\Big |_{{\rm UE}_N(x^{-1/2}(1 - x)^{1/2})} \nonumber \\
& = & E^{O^-(2N+1)}((0,\theta);\xi).
\end{eqnarray}
Here the average in the first equality is an example of the general form (\ref{2}) with
$g(x)$ having support on $(0,1)$ (an example of the Jacobi unitary ensemble JUE), while in
the second equality O${}^-(2N+1)$ denotes the ensemble of $(2N+1) \times (2N+1)$ random
orthogonal matrices with determinant $-1$, chosen with uniform (Haar) measure. The JUE generating
function is a $\tau$-function for a Painlev\'e VI system \cite{FW04}, thus evaluating
$E^{{\rm COE}_N+}$ and effectively solving the problem.

\subsection{The soft edge}
For the weight $g(x) = e^{-x^2}$ in (\ref{1}), corresponding to the Gaussian
unitary ensemble GUE, the support of the eigenvalue density is to leading order the
interval $(-\sqrt{2N}, \sqrt{2N})$. The largest eigenvalue is thus to leading order at 
$\sqrt{2N}$, but to higher order the eigenvalue density is nonzero to the right of this
point so it is referred to as the soft edge. Similarly for the weight $g(x) = x^a e^{-x}$ in
(\ref{1}) specifying the Laguerre unitary ensemble LUE, the leading support of the eigenvalue
density is $(0,4N)$, and the point $4N$ is referred to as the soft edge. It is
fundamental that, after scaling so that the eigenvalue spacing in the neighbourhood of the
respective soft edge is of order one, for $N \to \infty$ the gap probabilities at the
soft edge for both the GUE and LUE are identical, and given in terms of a Fredholm
determinant. Specifically, with
$$
E(J;{\rm ME}_N;\xi) := \sum_{n=0}^\infty (1 - \xi)^n E(n;J;{\rm ME}_N)
$$
one has \cite{Fo93a}
\begin{eqnarray}\label{6.1}
E_2^{\rm soft}((s,\infty);\xi) & := &
\lim_{N \to \infty} E((\sqrt{2N} + s/\sqrt{2} N^{1/6}, \infty); {\rm UE}_N(e^{-x^2});\xi)
\nonumber \\
& = & \lim_{N \to \infty} E((4N + 2 (2N)^{1/3}s, \infty); {\rm UE}_N(x^a e^{-x}); \xi)
\nonumber \\
& = & \det ( {\mathbb I} - \xi K_{(s,\infty)}^{\rm soft} )
\end{eqnarray}
(the superscript "2" in $E_2^{\rm soft}$ denotes the $\beta$ value of the symmetry type, here
unitary)
where $K^{\rm soft}_{(s,\infty)}$ is the integral operator on $(s,\infty)$ with kernel
\begin{eqnarray}\label{Ks}
 K^{\rm soft}(x,y) & = & {{\rm Ai}(x) {\rm Ai}'(y) - {\rm Ai}(y)  {\rm Ai}'(x) \over
x - y} \nonumber \\
 & = & \int_0^\infty {\rm Ai}(x+t)  {\rm Ai}(y+t)  \, dt,
\end{eqnarray}
Ai$(x)$ denoting the Airy function. It is similarly fundamental that $E_2^{\rm soft}$ admits
the Painlev\'e transcendent evaluation \cite{TW94a}
\begin{equation}\label{6.2}
E_2^{\rm soft}((s,\infty);\xi) = \exp \Big ( - \int_s^\infty (t - s) q^2(t;\xi) \, dt \Big )
\end{equation}
where $q(t)$ satisfies the particular Painlev\'e II equation
\begin{equation}
q'' = t q + 2q^3
\end{equation}
subject to the boundary condition
\begin{equation}\label{6.2a}
q(t;\xi) \mathop{\sim}\limits_{t \to \infty} \sqrt{\xi} {\rm Ai}(t).
\end{equation}

In the cases of orthogonal and symplectic symmetry, the soft edge scalings are specified by
\begin{eqnarray}\label{1s}
E_1^{\rm soft}((s,\infty);\xi)  & := & E((\sqrt{2N} + s/\sqrt{2} N^{1/6}, \infty);
{\rm OE}_N(e^{-x^2/2}); \xi) \nonumber \\
& := & E((4N + 2 (2N)^{1/3} s, \infty);
{\rm OE}_N(x^a e^{-x/2}); \xi)
\end{eqnarray}
and
\begin{eqnarray}\label{2s}
E_4^{\rm soft}((s,\infty);\xi)  & := & E((\sqrt{2N} + s/\sqrt{2} N^{1/6}, \infty);
{\rm SE}_{N/2}(e^{-x^2}); \xi) \nonumber \\
& := & E((4N + 2 (2N)^{1/3} s, \infty);
{\rm SE}_{N/2}(x^a e^{-x}); \xi).
\end{eqnarray}
Recently Dieng \cite{Di05a} has given the Painlev\'e transcendent evaluation
\begin{equation}\label{1.7}
\Big ( E_1^{\rm soft}((s,\infty);\xi) \Big )^2 =
E_2^{\rm soft}((s,\infty);\bar{\xi})
{(\xi - 1 - \cosh \mu(s,\bar{\xi}) + \sqrt{\bar{\xi}} \sinh \mu(s;\bar{\xi}) ) \over
\xi - 2}
\end{equation}
where
\begin{equation}
\bar{\xi} := 2 \xi - \xi^2, \qquad
\mu(s;\bar{\xi}) := \int_s^\infty q(t;\bar{\xi}) \, dt,
\end{equation}
together with
\begin{equation}\label{2.7}
\Big ( E_4^{\rm soft}((s,\infty);\xi)  \Big )^2 =
E_2^{\rm soft}((s,\infty);\xi) \cosh^2 {\mu(s;\xi) \over 2}.
\end{equation}
In the case $\xi=1$, (\ref{1.7}) and  (\ref{2.7}) reduce to results first derived by
Tracy and Widom \cite{TW96}. Indeed the study \cite{Di05a} uses the methods of
 \cite{TW96} to derive  (\ref{1.7}) and  (\ref{2.7}) for general $\xi$.

In the case $\xi=1$, the present author has provided simplified derivations of these
results. Moreover, in contrast to  \cite{TW96}, this simplified derivation makes essential
use of analogues of (\ref{C1}) and (\ref{C2}) for the Gaussian or Laguerre ensembles.
One of the objectives of this paper is to use this same approach to rederive
 (\ref{1.7}) and  (\ref{2.7}) for general $\xi$.
A satisfying feature of the derivation is that it offers an explanation for the peculiar
structure exhibited by (\ref{1.7}).

In the study \cite{DF06} it has been shown that
\begin{equation}\label{DF}
\exp \Big ( - \mu(s;\xi) \Big ) = {\det ( {\mathbb I} - \sqrt{\xi} V_{(0,\infty)}^{\rm soft})
\over \det ( {\mathbb I} + \sqrt{\xi} V_{(0,\infty)}^{\rm soft}) }
\end{equation}
where $ V_{(0,\infty)}^{\rm soft}$ is the integral operator on $(0,\infty)$ with kernel
$$
V^{\rm soft}(x,u) = {\rm Ai}(x+u+s).
$$
Note from (\ref{Ks}) that
$$
K_{(s,\infty)}^{\rm soft} \doteq (V_{(0,\infty)}^{\rm soft})^2,
$$
where here $\doteq$ means a change of variable is required in one of the integral
operators for equality
(such identities in random matrix theory are discussed at length
in the recent work \cite{Bl06}).
Further, it follows from (\ref{6.1}) that
\begin{equation}\label{DF1}
E_2^{\rm soft}((s,\infty);\xi) = \det ( {\mathbb I} - \sqrt{\xi} V_{(0,\infty)}^{\rm soft})
 \det ( {\mathbb I} + \sqrt{\xi} V_{(0,\infty)}^{\rm soft})
\end{equation}
(see \cite{DF06}). Consequently both (\ref{1.7}) and  (\ref{2.7}) can be expressed entirely in
terms of Fredholm determinants.

\begin{cor}\label{c.1}
One has
\begin{eqnarray}\label{DF2}
&&\Big ( E_1^{\rm soft}((s,\infty);\xi)  \Big )^2 = \Big ( {\xi - 1 \over \xi - 2} \Big )
\det ( {\mathbb I} - \sqrt{\xi} V_{(0,\infty)}^{\rm soft})
 \det ( {\mathbb I} + \sqrt{\xi} V_{(0,\infty)}^{\rm soft}) \nonumber \\
&& \qquad +
{\sqrt{\bar{\xi}} - 1 \over 2 (\xi - 2) }
\Big (  \det ( {\mathbb I} + \sqrt{\bar{\xi}} V_{(0,\infty)}^{\rm soft}) \Big )^2 -
{\sqrt{\bar{\xi}} + 1 \over 2 (\xi - 2) }
\Big (  \det ( {\mathbb I} - \sqrt{\bar{\xi}} V_{(0,\infty)}^{\rm soft}) \Big )^2 
\end{eqnarray}
and
\begin{equation}\label{DF3}
 E_4^{\rm soft}((s,\infty);\xi) = {1 \over 2} \Big (
 \det ( {\mathbb I} - \sqrt{{\xi}} V_{(0,\infty)}^{\rm soft})  +
 \det ( {\mathbb I} + \sqrt{{\xi}} V_{(0,\infty)}^{\rm soft}) \Big )
\end{equation}
\end{cor}

The results (\ref{DF1}) and (\ref{DF3}) should be compared against (\ref{1.7w}) and
(\ref{1.7x}). Note that when $\xi=1$, (\ref{DF2}) reduces to
\begin{equation}\label{DF4}
 E_1^{\rm soft}(0;(s,\infty)) = \det ( {\mathbb I} -  V_{(0,\infty)}^{\rm soft}).
\end{equation}
This is a known result, first conjectured by Sasamoto \cite{Sa05}, and subsequently proved by
Ferrari and Spohn \cite{FS05}.

\subsection{The hard edge}
The Laguerre ensembles have their origin in positive definite Hermitian matrices, and as
such all eigenvalues are positive. Because of this constraint the neighbourhood of the origin
in the Laguerre ensembles is referred to as the hard edge. This neighbourhood permits the
hard edge scaling limits \cite{Fo93a}
\begin{eqnarray}\label{1.24a}
E_1^{\rm hard}((0,s);a;\xi) & := & \lim_{N \to \infty}
E\Big ( \Big ( 0, {s \over 4N} \Big ); {\rm OE}_N(x^a e^{-x/2}) \Big ) \nonumber\\
E_2^{\rm hard}((0,s);a;\xi) & := & \lim_{N \to \infty}
E\Big ( \Big (0, {s \over 4N} \Big ); {\rm UE}_N(x^a e^{-x}) \Big ) \nonumber \\
E_4^{\rm hard}((0,s);a;\xi) & := & \lim_{N \to \infty}
E\Big ( \Big (0, {s \over 4N} \Big ); {\rm SE}_{N/2}(x^a e^{-x}) \Big ) 
\end{eqnarray}

Analogous to the results (\ref{6.1}) and (\ref{6.2}), for $\beta = 2$ one has the Fredholm
determinant evaluation \cite{Fo93a}
\begin{equation}\label{9.1}
E_2^{\rm hard}((0,s);a;\xi) =
 \det ( {\mathbb I} - \xi K_{(0,s)}^{\rm hard} )
\end{equation}
as well as the Painlev\'e transcendent evaluation \cite{TW94b}
\begin{equation}\label{9.2}
E_2^{\rm hard}((0,s);a;\xi) =
\exp \Big ( - {1 \over 4} \int_0^s \Big ( \log {s \over t} \Big ) \tilde{q}^2(t;a;\xi) \, dt
\Big ).
\end{equation}
In (\ref{9.1}) $ K_{(0,s)}^{\rm hard}$ is the integral operator on $(0,s)$ with kernel
\begin{eqnarray}\label{Kjh}
K^{\rm hard}(x,y) & = &
{J_a(\sqrt{x}) \sqrt{y} J_a'(\sqrt{y})  -  \sqrt{x}  J_a'(\sqrt{x}) J_a(\sqrt{y}) \over
x - y} \nonumber \\
& = &
{1 \over 4} \int_0^1 J_a(\sqrt{tx}) J_a(\sqrt{ty}) \, dt,
\end{eqnarray}
$J_a(z)$ denoting the Bessel function,
while in (\ref{9.2}) $\tilde{q}$ is the solution of the nonlinear equation
\begin{equation}
t ( \tilde{q}^2-1)(t\tilde{q}')' = \tilde{q} (t \tilde{q}')^2 + {1 \over 4} (t - a^2) \tilde{q}
+ {1 \over 4} \tilde{q}^3 (\tilde{q}^2 - 2)
\end{equation}
(a transformed version of the Painlev\'e V equation) subject to the boundary condition
\begin{equation}\label{1.29a}
 \tilde{q}(t;a;\xi) \mathop{\sim}\limits_{t \to 0^+} {\sqrt{\xi} \over 2^a \Gamma(1 + a) }
t^{a/2}.
\end{equation}

Let us consider now the $\beta = 1$ and 4 symmetries at the hard edge. Only the
$\xi = 1$ case of the following result was known previously \cite{Fo99b}.

\begin{thm}\label{t1}
One has
\begin{equation}\label{10.1}
\Big ( E_1^{\rm hard}((0,s);{a -1 \over 2};\xi) \Big )^2 =
E_2^{\rm hard}((0,s);a;\bar{\xi})
{(\xi - 1 - \cosh \tilde{\mu}(s;a;\bar{\xi}) + \sqrt{\bar{\xi}} \sinh 
\tilde{\mu}(s;a;\bar{\xi}) ) \over
\xi - 2}
\end{equation}
where
\begin{equation}\label{1.7a}
\bar{\xi} := 2 \xi - \xi^2, \qquad
\tilde{\mu}(s;a;\bar{\xi}) := {1 \over 2} \int_0^s \tilde{q}(t;a;\bar{\xi}) \, {dt \over \sqrt{t}},
\end{equation}
and
\begin{equation}\label{10.2}
\Big ( E_4^{\rm hard}((0,s);a+1;\xi)  \Big )^2 =
E_2^{\rm hard}((0,s);a;\xi) \cosh^2 {\tilde{\mu}(s;\xi) \over 2}.
\end{equation}
\end{thm}

Extending the working of \cite{Fo99b}, which relies crucially on the hard edge limit
of the Laguerre ensemble analogues of (\ref{C1}) and (\ref{C2}), Theorem \ref{t1}
will be established in Section 3.

We know from \cite{Fo93a, BF03} that in the limit $a \to \infty$, after appropriate scaling
the hard edge distributions reduce to the soft edge distributions. Explicitly, with
$$
a(\beta) = \left \{ \begin{array}{lll} (a-1)/2, & \beta = 1 \\
a, & \beta = 2 \\
a+1, & \beta = 4 \end{array} \right.
$$
we have
$$
\lim_{a \to \infty} E_\beta^{\rm hard}((0,a^2 - 2a(a/2)^{1/3} s); a(\beta);\xi) =
E_\beta^{\rm soft}((s,\infty);\xi).
$$
In the case $\xi = 1$ it was shown in \cite{Fo99b} that the soft edge Painlev\'e
transcendent evaluations (\ref{6.2}), (\ref{1.7}) and (\ref{2.7}), and those at the
hard edge (\ref{9.2}), (\ref{10.1}), (\ref{10.2}) are consistent with this result. The
same analysis applies for general $\xi$, showing that the hard edge result as
presented in  Theorem \ref{t1} contains (\ref{1.7}) and (\ref{2.7}) as a limiting case.

As with Corollary \ref{c.1}, recent results allow Theorem  \ref{t1} to be cast in terms of
Fredholm determinants. For this we need knowledge of the formulas \cite{DF06}
\begin{equation}
E_2^{\rm hard}((0,s);a;\xi) = \det ( {\mathbb I} - \sqrt{\xi} V^{\rm hard}_{(0,1)})
 \det ( {\mathbb I} + \sqrt{\xi} V^{\rm hard}_{(0,1)})
\end{equation}
and
\begin{equation}
\exp \Big ( - {1 \over 2} \int_0^s {\tilde{q}(t;a;\xi) \over \sqrt{t} } \, dt \Big )  =
{  \det ( {\mathbb I} - \sqrt{\xi} V^{\rm hard}_{(0,1)}) \over 
 \det ( {\mathbb I} + \sqrt{\xi} V^{\rm hard}_{(0,1)}) }.
\end{equation}
Here $ V^{\rm hard}_{(0,1)}$ is the integral operator on $(0,1)$ with kernel
\begin{equation}
V^{\rm hard}(x,y) = {\sqrt{s} \over 2} J_a(\sqrt{sxy}).
\end{equation}
Comparison with (\ref{Kjh}) shows
$$
K^{\rm hard}_{(0,s)} \doteq (V^{\rm hard}_{(0,1)})^2.
$$

\begin{cor}\label{c.2}
One has
\begin{eqnarray}\label{DF6}
&&\Big ( E_1^{\rm hard}((0,s);{a - 1 \over 2}; \xi)  \Big )^2 = \Big ( {\xi - 1 \over \xi - 2} \Big )
\det ( {\mathbb I} - \sqrt{\xi} V_{(0,1)}^{\rm hard})
 \det ( {\mathbb I} + \sqrt{\xi} V_{(0,1)}^{\rm hard}) \nonumber \\
&& \qquad +
{\sqrt{\bar{\xi}} - 1 \over 2 (\xi - 2) }
\Big (  \det ( {\mathbb I} + \sqrt{\bar{\xi}} V_{(0,1)}^{\rm hard}) \Big )^2 -
{\sqrt{\bar{\xi}} + 1 \over 2 (\xi - 2) }
\Big (  \det ( {\mathbb I} - \sqrt{\bar{\xi}} V_{(0,1)}^{\rm hard}) \Big )^2
\end{eqnarray}
and
\begin{equation}\label{DF7}
 E_4^{\rm hard}((s,\infty);a+1;\xi) = {1 \over 2} \Big (
 \det ( {\mathbb I} - \sqrt{{\xi}} V_{(0,1)}^{\rm hard})  +
 \det ( {\mathbb I} + \sqrt{{\xi}} V_{(0,1)}^{\rm hard}) \Big )
\end{equation}
\end{cor}

In the case $\xi = 1$ (\ref{DF6}) reduces to
\begin{equation}
 E_1^{\rm hard}(0;(0,s);{a - 1 \over 2}) 
= \det ( {\mathbb I} -  V_{(0,1)}^{\rm hard}),
\end{equation}
which is  known from \cite{DF06}.

\section{Soft edge}
\setcounter{equation}{0}
For the classical matrix ensembles of Hermitian matrices (classical weight functions
in (\ref{1}) and (\ref{3})) there are inter-relationships between the different symmetry
types analogous to (\ref{C1}) and (\ref{C2}). 
The existence of such identities were deduced indirectly in \cite{BR01a}. A direct study
and classification was undertaken in \cite{FR01}. 
Thus for the Gaussian ensembles one has
the superposition identity \cite{FR01}
\begin{equation}\label{B1}
{\rm even} \Big ( {\rm OE}_N(e^{-x^2/2}) \cup {\rm OE}_{N+1}(e^{-x^2/2}) \Big ) =
{\rm UE}_N(e^{-x^2}),
\end{equation}
as well as the decimation identity
\begin{equation}\label{B2}
{\rm even} \Big ( {\rm OE}_{2N+1}(e^{-x^2/2}) \Big ) = {\rm SE}_N(e^{-x^2}),
\end{equation}
where with
\begin{equation}\label{B3}
x_1 > x_2 > \cdots > x_{2N+1}
\end{equation}
the notation ``even" refers to the distribution of every even labelled eigenvalue.

These identities imply the following gap probability identities.

\begin{prop}\label{p1}
For $n \le N/2$
\begin{eqnarray}\label{A1}
&& E(n,(s,\infty);{\rm UE}_N(e^{-x^2}) ) 
= \sum_{p=0}^{2n+1} E(2n+1-p,(s,\infty);{\rm OE}_N( e^{-x^2/2})) \nonumber \\
&& \qquad \times
\Big ( E(p,(s,\infty);{\rm OE}_{N+1}(e^{-x^2/2})) +
E(p-1,(s,\infty);{\rm OE}_{N+1}(e^{-x^2/2})) \Big )
\end{eqnarray}
and for $n \le N$
\begin{equation}\label{A2}
 E(n,(s,\infty);{\rm SE}_N(e^{-x^2}) ) =
 E(2n,(s,\infty);{\rm OE}_{2N+1}(e^{-x^2/2 }) ) +
E(2n+1,(s,\infty);{\rm OE}_{2N+1}(e^{-x^2/2 }) ) 
\end{equation}
where $E(-1,\cdot) := 0$.
\end{prop}

\noindent 
{\it Proof} \quad Consider first (\ref{A1}). From the superposition identity (\ref{B1}), the
ensemble UE${}_N(e^{-x^2})$ results by integrating out the even labelled
eigenvalues in the superimposed ensembles. It then follows from (\ref{B3}) that to have
exactly $n$ eigenvalues in the interval $(s,\infty)$, the superimposed ensemble must
contain either $2n$ or $2n+1$ eigenvalues. Given a partition
$2n  = (2n+1-p) + (p-1)$ such that $2n+1-p$ eigenvalues are from OE${}_N(e^{-x^2})$ and $p-1$
eigenvalues are from OE${}_{N+1}(e^{-x^2})$, it follows from (\ref{B1}) that the
probability  UE${}_N(e^{-x^2})$ has $n$ eigenvalues in $(s,\infty)$ is
\begin{equation}\label{A3}
E(2n+1-p,(s,\infty);{\rm OE}_N(e^{-x^2/2}))
E(p-1,(s,\infty);{\rm OE}_{N+1}(e^{-x^2/2})).
\end{equation}
Similarly, given a partition $2n+1 = (2n+1-p) + p$ of the $2n+1$ superimposed eigenvalues
in $(s,\infty)$, such that $(2n+1-p)$ eigenvalues are from  OE${}_N(e^{-x^2})$ and $p$
eigenvalues are from OE${}_{N+1}(e^{-x^2})$, the probability that UE${}_N(e^{-x^2})$ has
$n$ eigenvalues in $(s,\infty)$ is
\begin{equation}\label{A4}
E(2n+1-p,(s,\infty);{\rm OE}_N(e^{-x^2/2}))
E(p,(s,\infty);{\rm OE}_{N+1}(e^{-x^2/2})).
\end{equation}
Summing (\ref{A3}) and (\ref{A4}) over $p$ gives (\ref{A1}). 

To derive (\ref{A2}), simply note from (\ref{B2}) and (\ref{B3}) that for SE${}_N(e^{-x^2})$
to have exactly $n$ eigenvalues in $(s,\infty)$, OE${}_{2N+1}(e^{-x^2/2})$ must have
either $2n$ or $2n+1$ eigenvalues in $(s,\infty)$. \hfill $\square$

\medskip
Using the definitions of the soft edge scalings in (\ref{6.1}), (\ref{1s}) and
(\ref{2s}), by replacing $s \mapsto \sqrt{2N} + s/\sqrt{2} N^{1/6}$ in (\ref{A1})
and (\ref{A2}), and taking the $N \to \infty$ limit, we obtain the following soft edge
inter-relations.

\begin{cor}
One has
\begin{equation}\label{aD}
E_2^{\rm soft}(n,(s,\infty)) = \sum_{p=0}^{2n+1} E_1^{\rm soft}(2n+1-p,(s,\infty))
\Big ( E_1^{\rm soft}(p,(s,\infty)) + E_1^{\rm soft}(p-1,(s,\infty)) \Big )
\end{equation}
and
\begin{equation}\label{bD}
E_4^{\rm soft}(n,(s,\infty)) = E_1^{\rm soft}(2n+1,(s,\infty)) +
 E_1^{\rm soft}(2n,(s,\infty)). 
\end{equation}
Equivalently, with $\bar{\xi}$ specified in (\ref{1.7a}), in terms of generating
functions
\begin{equation}\label{aE}
(1 - \xi) E_2^{\rm soft}((s,\infty);\bar{\xi}) =
\mathop{\rm odd}\limits_{1 - \xi} \Big [ \Big (E_1^{\rm soft}((s,\infty);\xi) \Big )^2
(2 - \xi) \Big ]
\end{equation}
and
\begin{equation}\label{bE}
(1 - \xi) E_4^{\rm soft}((s,\infty);\bar{\xi}) =
\mathop{\rm odd}\limits_{1 - \xi} \Big [ E_1^{\rm soft}((s,\infty);\xi) 
(2 - \xi) \Big ]
\end{equation}
where the notation $\mathop{\rm odd}\limits_{1 - \xi}\, f$ denotes the odd powers in
$1 - \xi$ of the expansion of $f(\xi)$ about $\xi=1$.
\end{cor}

\noindent
{\it Proof.} \quad It remains to deduce the generating function identities (\ref{aE}),
(\ref{bE}) from (\ref{aD}), (\ref{bD}). For this multiply both sides by
$(1 - \xi)(1 - \bar{\xi})^n$ and sum over $n$, noting on the right hand side that
$(1 -  \bar{\xi}) = (1 - \xi)^2$, and in (\ref{aD}) making use of the formula for the
multiplication of two power series. \hfill $\square$

\medskip
Comparing (\ref{aE}) with (\ref{1.7}) it must be that
\begin{equation}\label{cE}
E_2^{\rm soft}((s,\infty);\bar{\xi}) 
\Big ( \cosh \mu(s,\bar{\xi}) - \sqrt{\bar{\xi}} \sinh \mu(s,\bar{\xi}) \Big )
=
\mathop{\rm even}\limits_{1 - \xi} \Big [ \Big (E_1^{\rm soft}((s,\infty);\xi) \Big )^2
(2 - \xi) \Big ].
\end{equation}
Conversely, (\ref{cE}) and (\ref{aE}) together imply (\ref{1.7}). Our remaining task is thus
to derive (\ref{cE}).

For this purpose, following \cite{Fo99b}, we consider the soft edge scaling limit of the ensemble
$$
{\rm odd} \Big ( {\rm OE}_N(e^{-x^2}) \cup {\rm OE}_N(e^{-x^2}) \Big )
$$
where the notation ``odd" refers to the distribution of every odd labelled eigenvalue
in the superimposed ensemble.
As shown in \cite{Fo99b}, the $n$-point correlation function is given by 
\begin{eqnarray}\label{7.crag}
\lefteqn{
\rho_{(n)}^{{\rm odd (OEsoft)}^2}(x_1,\dots,x_n)} \nonumber \\
 && :=
\lim_{N \to \infty} \Big ( {1 \over 2^{1/2} N^{1/6}} \Big )^n
\rho_{(n)}^{{\rm odd( GOE}_N)^2} \Big (
\sqrt{2N} + {x_1 \over 2^{1/2} N^{1/6}}, \dots,
\sqrt{2N} + {x_n \over 2^{1/2} N^{1/6}} \Big ) \nonumber \\
& & = \det \Big [
K^{\rm soft}(x_j,x_k) + {\rm Ai}(x_j) \int_0^\infty {\rm Ai}(x_k - v) \,
dv \Big ]_{j,k=1,\dots,n}
\end{eqnarray}
Consequently, the generating function for the gap probabilities in $(s,\infty)$ is given by
the Fredholm determinant formula
\begin{equation}\label{AB}
E^{{\rm odd (OEsoft)}^2}((s,\infty);\xi) =
\det \Big ( 1 - \xi ( K^{\rm soft}_{(s,\infty)} + A^{\rm s} \otimes B^{\rm s}) \Big )
\end{equation}
where $K^{\rm soft}_{(s,\infty)}$ is as in  
(\ref{6.1}), $A^{\rm s}$ is the operator which multiplies by ${\rm Ai}(x)$, and
$B^{\rm s}$ is the integral operator on $(s,\infty)$ with kernel $\int_0^\infty
{\rm Ai}(y-v) \, dv$.

Analogous to the result (\ref{aD}) following from the superposition identity (\ref{B1}), the
ensemble ${\rm odd (OEsoft)}^2$ being constructed as a superposition implies the following
gap probabilites identities.

\begin{prop}\label{p2}
One has
\begin{equation}\label{a0}
E^{{\rm odd (OEsoft)}^2}(n;(s,\infty)) = \sum_{p=0}^{2n} E_1^{\rm soft}(2n-p,(s,\infty))
\Big ( E_1^{\rm soft}(p,(s,\infty)) + E_1^{\rm soft}(p-1,(s,\infty)) \Big ).
\end{equation}
Equivalently, with $\bar{\xi}$ specified in (\ref{1.7a}), in terms of generating
functions
\begin{equation}\label{ae}
 E^{{\rm odd (OEsoft)}^2}((s,\infty);\bar{\xi}) =
\mathop{\rm even}\limits_{1 - \xi} \Big [ \Big (E_1^{\rm soft}((s,\infty);\xi) \Big )^2
(2 - \xi) \Big ]
\end{equation}
(cf.~(\ref{aE})).
\end{prop}

We now seek to independently evaluate $ E^{{\rm odd (OEsoft)}^2}((s,\infty);{\xi})$.
In \cite{Fo99b} it was shown from the Fredholm determinant formula (\ref{AB}) that
\begin{equation}\label{x1}
E^{{\rm odd (OEsoft)}^2}((s,\infty);\xi) \Big |_{\xi = 1} =
\bigg (  E_2^{\rm soft}((s,\infty);\xi) \exp \Big ( - \int_s^\infty q(t;\xi) \, dt
\Big ) \bigg ) \bigg |_{\xi = 1}.
\end{equation}
From the inter-relation (\ref{a0}), and the definition of the generating functions, this is
equivalent to the evaluation \cite{TW96}
$$
\Big ( E_1^{\rm soft}(0;(s,\infty)) \Big )^2 =
  E_2^{\rm soft}(0;(s,\infty)) \exp \Big ( - \int_s^\infty q(t;1) \, dt \Big ) .
$$
The $\xi = 1$ result (\ref{x1}) can readily be generalized to arbitrary $\xi$.

\begin{prop}
Let $q(t;\xi)$ be as in (\ref{6.2}), and thus having its $\xi$ dependence determined by the
boundary condition (\ref{6.2a}). Let $\mu(s;\xi)$ be as in (\ref{1.7a}). We have
\begin{equation}\label{p.1}
E^{{\rm odd (OEsoft)}^2}((s,\infty);\xi) =
 E_2^{\rm soft}((s,\infty);\xi) \Big ( \cosh \mu(s;\xi) - \sqrt{\xi}
\sinh \mu(s;\xi) \Big ).
\end{equation}
\end{prop}

\noindent
{\it Proof. } \quad We see from (\ref{AB}) that
\begin{equation}\label{p.2}
E^{{\rm odd (OEsoft)}^2}((s,\infty);\xi) 
 = \det ( \mathbb I - \xi K^{\rm soft}_{(s,\infty)})
\Big ( 1 - \xi \int_s^\infty [(\mathbb I
 - \xi K^{\rm soft}_{(s,\infty)})^{-1}A^{\rm s}](y) B^{\rm s}(y) \, dy \Big )
\end{equation} 
where use has been made of the fact that the direct product of two operators has rank 1. 
Recalling (\ref{6.1}), we see that (\ref{p.1}) becomes equivalent to showing
\begin{equation}\label{p.3}
 1 - \xi \int_s^\infty [(1 - \xi K^{\rm soft}_{(s,\infty)})^{-1}A^{\rm s}](y) B^{\rm s}(y) \, dy =
\cosh \mu(s;\xi) - \sqrt{\xi} \sinh \mu(s;\xi).
\end{equation}
For this we closely follow \cite{Fo99b}, where this result for $\xi = 1$ was derived.

In terms of the notation
$$
\phi^{\rm s}(x) := \sqrt{\xi} {\rm Ai}(x), \qquad
Q^{\rm s}(x) := [(1 - \xi K^{\rm soft})^{-1} \phi^{\rm s}](x)
$$
we have that
\begin{equation}\label{4.14}
\xi \int_s^\infty  [(1 - \xi K^{\rm soft}_{(s,\infty)})^{-1}A^{\rm s}](y) B^{\rm s}(y) \, dy =
\int_s^\infty dy \, Q^{\rm s}(y) \int_{-\infty}^y \phi^{\rm s}(v) \, dv =: u_\epsilon^{\rm s}.
\end{equation}
Now introduce the quantity
\begin{equation}\label{4.15}
q_\epsilon^{\rm s} := \int_s^\infty dy \, \rho^{\rm s}(s,y) \int_{-\infty}^y \phi^{\rm s}(v) \, dv,
\end{equation}
where $\rho^{\rm s}(s,y)$ denotes the kernel of the integral operator 
$(\mathbb I - \xi K^{\rm soft}_{(s,\infty)})^{-1}$.
According to the working of \cite[eqs.~(3.11)--(3.14)]{Fo99b} the quantities $u_\epsilon^{\rm s}$
and $q_\epsilon^{\rm s}$ satisfy the coupled equations
\begin{eqnarray}
{d u_\epsilon^{\rm s} \over ds} & = & - q(s;\xi) q_\epsilon^{\rm s} \label{4.16} \\
{d q_\epsilon^{\rm s} \over ds} & = &  q(s;\xi) (1 - u_\epsilon^{\rm s}) \label{4.17}
\end{eqnarray}
where $q(s;\xi)$ enters via the fact that $Q^{\rm s}(s) = q(s;\xi)$.

Since $Q^{\rm s}(y)$ is smooth while $\rho^{\rm s}(s,y)$ is equal to the delta function $\delta(s-y)$ plus a
smooth term, we see from (\ref{4.14}), (\ref{4.15}), and the fact that
$$
\int_{-\infty}^\infty \phi^{\rm s}(v) \, dv = \sqrt{\xi},
$$
that (\ref{4.16}), (\ref{4.17}) must be solved subject to the boundary conditions
\begin{equation}\label{2.24a}
u_\epsilon^{\rm s} \to 0, \qquad q_\epsilon^{\rm s} \to \sqrt{\xi} \quad {\rm as} \quad s \to \infty.
\end{equation}
It is straightforward to verify that the solution subject to these boundary conditions is
\begin{eqnarray*}
&&  q_\epsilon^{\rm s} = \sqrt{\xi} \cosh \mu(s;\xi) - \sinh \mu(s;\xi) \\
&& u_\epsilon^{\rm s} = 1 -  \cosh \mu(s;\xi) + \sqrt{\xi} \sinh \mu(s;\xi).
\end{eqnarray*}
The latter is equivalent to (\ref{p.3}), so completing our derivation. \hfill $\square$

\medskip
Substituting (\ref{p.1}) in (\ref{aE}) we obtain (\ref{cE}). As already noted, this together with
(\ref{aE}) implies (\ref{1.7}).

We turn our attention now to (\ref{2.7}). Dieng \cite{Di05a} has provided an elementary proof that
(\ref{1.7}) and (\ref{2.7}) together imply (\ref{bD}). Since (\ref{bD}) uniquely determines
$E_4^{\rm soft}((s,\infty);\xi)$ in terms of $\{E_1^{\rm soft}(n;(s,\infty)) \}_{n=0,1,\dots}$
it follows then that (\ref{1.7}) and (\ref{bD}) imply (\ref{2.7}).

\section{Hard edge}
\setcounter{equation}{0}
The key formulas of the previous section relating to the soft edge all have hard edge
analogues. 
Noting from the definitions
(\ref{1.24a}) of the hard edge scaling limit, we first make note of inter-relationships between
Laguerre ensembles of different symmetry types. According to \cite{FR01}, one has the
superposition identity
\begin{equation}\label{2.37h}
{\rm even} \Big ( {\rm OE}_N(x^{(a-1)/2}
e^{-x/2}) \cup {\rm OE}_{N+1}(x^{(a-1)/2} e^{-x/2}) \Big ) =
{\rm UE}_N(x^a e^{-x}),
\end{equation}
as well as the decimation identity
\begin{equation}\label{2.38h}
{\rm even} \Big ( {\rm OE}_{2N+1}(x^{(a-1)/2} e^{-x/2}) \Big ) = {\rm SE}_N(x^{a+1} e^{-x}).
\end{equation}
Arguing as in the derivation of Proposition \ref{p1} shows that these identities imply the
following gap probability identities.

\begin{prop}
For $n \le N/2$
\begin{eqnarray}\label{A1h}
&& E(n,(0,s);{\rm UE}_N(x^a e^{-x}) )
= \sum_{p=0}^{2n+1} E(2n+1-p,(0,s);{\rm OE}_N( x^{(a-1)/2} e^{-x/2})) \nonumber \\
&& \qquad \times
\Big ( E(p,(0,s);{\rm OE}_{N+1}(x^{(a-1)/2} e^{-x/2})) +
E(p-1,(0,s);{\rm OE}_{N+1}(x^{(a-1)/2} e^{-x/2})) \Big )
\end{eqnarray}
and for $n \le N$
\begin{eqnarray}\label{A2h}
&& E(n,(0,s);{\rm SE}_N(x^{a+1} e^{-x}) ) \nonumber \\
&& \quad =
 E(2n,(0,s);{\rm OE}_{2N+1}(x^{(a-1)/2} e^{-x/2 }) ) +
E(2n+1,(0,s);{\rm OE}_{2N+1}(x^{(a-1)/2} e^{-x/2 }) ).
\end{eqnarray}
\end{prop}

By replacing $s \mapsto s/4N$, and recalling (\ref{1.24a}), the hard edge scaling limit of these
gap probabilities can be taken immediately.

\begin{cor}
One has
\begin{eqnarray}\label{aDh}
E_2^{\rm hard}(n,(0,s);a) & = & \sum_{p=0}^{2n+1} E_1^{\rm hard}(2n+1-p,(0,s);(a-1)/2)
\nonumber \\
&& \times
\Big ( E_1^{\rm hard}(p,(0,s);(a-1)/2) + E_1^{\rm hard}(p-1,(0,s);(a-1)/2) \Big )
\end{eqnarray}
and
\begin{equation}\label{bDh}
E_4^{\rm hard}(n,(0,s);a+1) = E_1^{\rm hard}(2n+1,(0,s);(a-1)/2) +
 E_1^{\rm hard}(2n,(0,s);(a-1)/2).
\end{equation}
The first is equivalent to the generating function identity
\begin{equation}\label{aEh}
(1 - \xi) E_2^{\rm hard}((0,s);a;\bar{\xi}) =
\mathop{\rm odd}\limits_{1 - \xi} \Big [ \Big (E_1^{\rm hard}((0,s);(a-1)/2;\xi) \Big )^2
(2 - \xi) \Big ].
\end{equation}
\end{cor}

With (\ref{aEh}) established, we see that (\ref{10.1}) is equivalent to the identity
\begin{equation}\label{cEh}
E_2^{\rm hard}(((0,s);a;\bar{\xi})
\Big ( \cosh \tilde{\mu}(s,a,\bar{\xi}) - \sqrt{\bar{\xi}} \sinh \tilde{\mu}(s,a,\bar{\xi}) \Big )
=
\mathop{\rm even}\limits_{1 - \xi} \Big [ \Big (E_1^{\rm hard}((0,s);a;\xi) \Big )^2
(2 - \xi) \Big ]
\end{equation}
(cf.~(\ref{cE})).

To derive (\ref{cEh}), we follow \cite{Fo99b} and consider the hard edge scaling of the
ensemble
$$
{\rm even} \Big ( {\rm OE}_N(x^{(a-1)/2} e^{-x/2}) \cup {\rm OE}_N(x^{(a-1)/2} e^{-x/2}) \Big )
=: {\rm even} ({\rm LOE}_N)^2.
$$
It is shown in \cite{Fo99b} that the corresponding $n$-point correlation is
\begin{eqnarray}\label{7.cral}
\rho_{(n)}^{{\rm odd(OEhard)}^2}(x_1,\dots,x_n) & :=
& \lim_{N \to \infty} \Big ( {1 \over 4N} \Big )^n
\rho_{(n)}^{{\rm even (LOE}_N)^2} \Big (
{x_1 \over 4N}, \dots, {x_n \over 4N} \Big ) \nonumber \\
& = & \det \Big [
K^{\rm hard}(x_j,x_k) + {J_a(\sqrt{x_j}) \over 2 \sqrt{x_k}}
\int_{\sqrt{x_k}}^\infty J_a(t) \, dt \Big ]_{j,k=1,\dots,n},
\end{eqnarray}
where the notation ${\rm odd(OEhard)}^2$ on the left hand side refers to the ordering
$x_1 < x_2 < \cdots$ which is appropriate at the hard edge in the scaling limit. We see from this that
\begin{equation}\label{3.10}
E^{{\rm odd(OEhard)}^2}((0,s);\xi;(a-1)/2) = \det \Big ( 1 - \xi ( K^{\rm hard}_{(0,s)} + A^{\rm h}
\otimes B^{\rm h} ) \Big )
\end{equation}
where $K^{\rm hard}_{(0,s)}$ is as in (\ref{9.1}), while $ A^{\rm h}$ is the operator which
multiplies by $J_a(\sqrt{x})$ and $B^{\rm h}$ is the integral operator on $(0,s)$
with kernel ${1 \over 2 \sqrt{y}} \int_y^\infty J_a(t) \, dt$.

The significance of the ensemble ${\rm odd(OEhard)}^2$ for the problem at hand is that
its gap probabilities are, from it definition as a superposition, given in terms of the
$\beta = 1$ hard edge gap probabilities.

\begin{prop}\label{p2h}
One has
\begin{eqnarray}\label{a0h}
&& E^{{\rm odd (OEhard)}^2}(n;(0,s);(a-1)/2)  = \sum_{p=0}^{2n} E_1^{\rm hard}(2n-p,(0,s);(a-1)/2)
\nonumber \\
&& \qquad \times
\Big ( E_1^{\rm hard}(p,(0,s);(a-1)/2) + E_1^{\rm hard}(p-1,(0,s);(a-1)/2) \Big ),
\end{eqnarray}
or equivalently, in terms of generating functions, 
\begin{equation}\label{aeh}
 E^{{\rm odd (OEhard)}^2}((0,s);(a-1)/2;\bar{\xi}) =
\mathop{\rm even}\limits_{1 - \xi} \Big [ \Big (E_1^{\rm hard}((0,s);(a-1)/2;\xi) \Big )^2
(2 - \xi) \Big ]
\end{equation}
where $\bar{\xi}$ is as in (\ref{9.2}).
\end{prop}

We see from (\ref{aeh}) that to  derive (\ref{cEh}) we need to evaluate $E^{{\rm odd (OEhard)}^2}$.
In \cite{Fo99b} it was shown that
\begin{equation}\label{x1h}
E^{{\rm odd (OEhard)}^2}((0,s);\xi;(a-1)/2) \Big |_{\xi = 1} =
\bigg (  E_2^{\rm hard}((0,s);\xi;a) \exp \Big ( - {1 \over 2}
\int_0^s {\tilde{q}(t;a;\xi) \over \sqrt{t} }\, dt
\Big ) \bigg ) \bigg |_{\xi = 1}.
\end{equation}
As in going from (\ref{a0}) to (\ref{x1}), this can readily be generalized to arbitrary $\xi$.

\begin{prop}
Let $\tilde{q}(t;a;\xi)$ be as in (\ref{9.2}), and thus having its $\xi$ dependence determined by the
boundary condition (\ref{1.29a}). Let $\tilde{\mu}(s;a;\xi)$ be as in (\ref{1.7a}). We have
\begin{equation}\label{p.1h}
E^{{\rm odd (OEhard)}^2}((0,s);\xi;(a-1)/2) =
 E_2^{\rm hard}((0,s);\xi;a) \Big ( \cosh \tilde{\mu}(s;\xi;a) - \sqrt{\xi}
\sinh \tilde{\mu}(s;\xi;a) \Big ).
\end{equation}
\end{prop}

\noindent
{\it Proof. } \quad As in going from (\ref{p.2}) to (\ref{p.3}), it follows from  (\ref{3.10})
that (\ref{p.1h}) is equivalent to the identity
\begin{equation}\label{p.3h}
 1 - \xi \int_0^s [({\mathbb I} - \xi K^{\rm hard}_{(0,s)})^{-1}A^{\rm h}](y) B^{\rm h}(y) \, dy =
\cosh \tilde{\mu}(s;\xi;a) - \sqrt{\xi} \sinh \tilde{\mu}(s;\xi;a).
\end{equation}
This was derived in \cite{Fo99b} for $\xi=1$, and we adopt the same approach for the general $\xi$
case. Setting
$$
\phi^{\rm h}(x) = \sqrt{\xi} J_a(\sqrt{x}), \qquad
Q^{\rm h}(x) = [ ({\mathbb I} - \xi K^{\rm hard})^{-1} \phi^{\rm h} ](x)
$$
and after changing variables $t = \sqrt{u}$ in the definition of $B^{\rm h}$ we have
$$
\xi \int_0^s [({\mathbb I} - \xi K^{\rm hard}_{(0,s)})^{-1}A^{\rm h}](y) B^{\rm h}(y) \, dy =
{1 \over 4} \int_0^s dy \, Q^{\rm h}(y) {1 \over \sqrt{y} }
\int_y^\infty du \, {1 \over \sqrt{u} } \phi^{\rm h}(u) =: u_\epsilon^{\rm h}.
$$
Also introduce 
$$
q_\epsilon^{\rm h} := \int_0^s dy \, \rho^{\rm h}(s,y)
\int_{-\infty}^y du \, {1 \over \sqrt{u} } \phi^{\rm h}(u)
$$
where $ \rho^{\rm h}(s,y)$ denotes the kernel of the integral operator
$(1 - \xi K^{\rm hard}_{(0,s)})^{-1}$. According to the working of
\cite[eqns.~(3.23)--(3.27)]{Fo99b}, with $\tilde{q}_\epsilon^{\rm h} := \sqrt{s}
q_\epsilon^{\rm h}$ the quantities $ u_\epsilon^{\rm h}$ and $\tilde{q}_\epsilon^{\rm h}$
satisfy the coupled system of equations
\begin{eqnarray*}
\sqrt{s} {d u_\epsilon^{\rm h} \over ds} & = & {1 \over 4}
\tilde{q}(s;a;\xi) \tilde{q}_\epsilon^{\rm h} \\
\sqrt{s} {d \tilde{q}_\epsilon^{\rm h} \over ds} & = & - 
\tilde{q}(s;a;\xi) (1 - u_\epsilon^{\rm h}).
\end{eqnarray*}
Arguing as in the derivation of (\ref{2.24a}) shows that these equations must be solved
subject to the boundary conditions
$$
 u_\epsilon \to 0, \qquad \tilde{q}_\epsilon^{\rm h} \to 2 \sqrt{\xi}
\quad {\rm as} \quad s \to \infty.
$$
We verify that the solution subject to these boundary conditions is
\begin{eqnarray*}
&&\tilde{q}_\epsilon^{\rm h} = 2 \Big (
\sqrt{\xi} \cosh \tilde{\mu}(s;a;\xi) - \sinh \tilde{\mu}(s;a;\xi) \Big )\\
&& u_\epsilon^{\rm h} = 1 -  \cosh \tilde{\mu}(s;a;\xi) + \sqrt{\xi} \sinh \tilde{\mu}(s;a;\xi).
\end{eqnarray*}
The latter is equivalent to (\ref{p.3h}), so our proof is complete.
\hfill $\square$

\medskip
Substituting (\ref{p.1h}) in (\ref{aeh}) establishes (\ref{cEh}), and so completes our derivation
of (\ref{10.1}). Because the structure of (\ref{10.1}) is identical to that of the soft edge
formula (\ref{1.7}), we can appeal to Dieng's proof \cite{Di05a} that (\ref{10.1}) and
(\ref{2.7}) together imply (\ref{bD}), which does not make use of the explicit form of
$\mu(s;\xi)$ in (\ref{1.7}), to conclude that  (\ref{10.1}) and (\ref{10.2}) imply
(\ref{bDh}). But (\ref{bDh}) uniquely determines $E_4^{\rm hard}((0,s);\xi;a+1)$ in terms of
$\{E_1^{\rm soft}(n;(0,s);(a-1)/2)\}_{n=0,1,\dots}$ so it follows that  (\ref{10.1}) and
(\ref{bDh}) imply  (\ref{10.2}).

\subsection*{Acknowledgements}
This work was supported by the Australian Research Council.


\end{document}